\documentclass[
reprint,
 preprintnumbers,
nofootinbib,
 amsmath,amssymb,
 aip,
]{revtex4-2}
\usepackage{hyperref}
\hypersetup{
    colorlinks=true,
    linkcolor=black,
    filecolor=magenta,      
    urlcolor=blue,
    citecolor=cyan,
    }
\usepackage{graphicx}% Include figure files
\usepackage{dcolumn}% Align table columns on decimal point
\usepackage{bm}% bold math
\usepackage{color}
\usepackage[color,matrix,arrow]{xy}

\usepackage{color}

\usepackage{xr} % To refer figures and equation in supplementary information
\usepackage{amsmath}
\usepackage{cleveref}
\usepackage{natbib}

\begin{document}

\title{Factors that control stability, variability, and reliability issues of endurance cycle in ReRAM devices: a phase field study}

%\homepage[\copyright 2022. This version of the manuscript is made available under the CC-BY-NC-ND 4.0 license.]{ }

%\author{Arijit Roy\corref{mycorrespondingauthor}}
%\address{Deoartment of Materials Science and Engineering, Kookmin University, Seoul 02707, Republic of Korea}
%\ead{arijitroy@kookmin.ac.kr}

%\author{Pil-Ryung Cha\corref{mycorrespondingauthor}}
%\address{Department of Materials Science and Engineering, Kookmin University, Seoul 02707, Republic of Korea}
%\ead{cprdream@kookmin.ac.kr}

\author{Arijit Roy}%
\email{arijitroy@kookmin.ac.kr}
\affiliation{School of Materials Science and Engineering, Kookmin University, Seoul 02707, Republic of Korea}%

\author{Min-Gyu Cho}%
\affiliation{School of Materials Science and Engineering, Kookmin University, Seoul 02707, Republic of Korea}%

\author{Pil-Ryung Cha}
 \email{cprdream@kookmin.ac.kr}
\affiliation{School of Materials Science and Engineering, Kookmin University, Seoul 02707, Republic of Korea}%


\begin{abstract} 
The morphological evolution of the conducting filament (CF) predominantly controls the electric response of the resistive random access memory (ReRAM) devices. However, the parameters -- in terms of the material and the processing -- which control the growth of such CF are plenty. Extending the phase field technique for ReRAM systems presented by Roy and Cha [\href{https://doi.org/10.1063/5.0026350}{J. Appl. Phys. 128, 205102 (2020)}], we could successfully model the complete SET (low resistance state) and RESET (high resistance state) sates due to the application of sweeping voltage. The key parameters that influence the stability of the multi-cycle \emph{I-V} response or the endurance behavior are identified. The computational findings of the presented model ReRAM system are practical in correlating the multi-parametric influence with the stability, variability, and reliability of the endurance cycle that affect the device performance and also lead to the device failure. We believe that our computational approach of connecting the morphological changes of the CF with the electrical response, has the potential to further understand and optimize the performance of the ReRAM devices.
\end{abstract}


\maketitle

\section{Introduction}
The development of the memresistive devices are crucial for the advancement of electronics of the future. Memresistive devices are poised to tackle the needs of future electronics in a variety of challenging areas of integrated circuit designs and non-volatile memories. The most common application of memresistive devices are in the field of resistive Random Access Memories (ReRAM). ReRAM has already proved to be effective in terms of scalability, fabrication complexity, latency, and low power consumption as compared to its predecessor, i.e. static RAM (SRAM), dynamic RAM (DRAM), and flash memories~\cite{Zahoor2020}. Also, the reduced fabrication complexity of simple metal--insulator--metal (MIM) structure and compatibility with existing CMOS (complementary metal oxide semiconductor) field effect transistor (FET) technology provides ReRAM an edge as compared to its other competitors; phase change memory (PCM) cells and spin-transfer torque magnetoresistive random access memories (STT-MRAM)~\cite{Chappert2009}. Other relevant areas of application of the memresistive devices include the reconfigurable logic circuits~\cite{Strukov2005,Xia2009,Tan2017} and the brain inspired neuromorphic computing devices~\cite{Yang2015,Burr2017,VanDeBurgt2017,Guo2020}. Implementation of logic and memory in same circuits are realized in the reconfigurable logic circuits as an alternative to meet the ever increasing demand of scalability which is governed by the Moore's law~\cite{Moore1965}. On the other hand to realize the electronic devices consisting of artificial synapses to harness the power of data driven learning and decision making capabilities, memresistive devices are explored and optimised continuously. 

In this paper, we have looked into the aspect of understanding the morphological changes that happen inside the ReRAM systems during the application of electrical stimuli. More specifically, we are interested in connecting the electrical responses due to the morphological changes inside the memresistive systems and predict the device performance and failure mechanism. However, due to the presence of multi-parametric dependence of material properties, it is challenging to understand the factors that affect the device performance and predict the exact cause of device failure. To eliminate such challenges in order to improve the device performance, it is important to understand the materials and processing conditions that affect it and accordingly choosing the class of compatible materials which are suitable for the use in resistive switching devices. 
 
The operation mechanism of memresistive devices are as simple as its device architecture, which mainly consists of MIM structure. The electrodes are responsible for maintaining the biasing potential and are also responsible in supplying the charged species to promote the electrical responses in case of electrochemical metallization memory (ECM) cells. The sandwiched memresistive layer is responsible for controlling the thermodynamics and growth kinetics of the conducting filament (CF) phase on the counter electrode due to the supplied ionic species from the active electrode. During the growth of CF due to the applied sweeping bias, the magnitude of measured current remains low and increases steadily. The state of the resistance at this stage is called high resistance state or HRS. With the increasing magnitude of applied voltage sweeping, at a particular voltage the growth of CF completes and thereby short-circuit the electrodes, leading to an abrupt increase of measured current in the device. The process of such formation of CF inside the memresistive layer is called forming and the voltage at which forming of CF completes is called forming voltage. The state of the resistance at this stage of operation is called low resistance state or LRS~\cite{Roy2018,Roy2020}. By reversing the direction of voltage sweeping cycle, it is possible that for a specific applied bias, the filament breaks and as a consequence the resistivity of the device increases. This process is called RESET. Again by reversing the direction of voltage sweeping cycle -- during SET process -- the filament formation completes and thereby the HRS state is converted back to the LRS state. Such repeating cycles -- also called endurance cycle -- of filament completion and rupture define the specific electrical characteristics of the memristor devices.  

Implementation challenges of memristor devices for ReRAM related applications are of many folds. One of the prominent reason is the challenge related to the reliability in terms of electrical responses during the endurance cycle. The process of repeatedly forming and rupturing the CF are governed by the following key physio-chemical processes: a) migration of charged species inside the memresistive layer due to the applied electric potential, b) formation of CF nucleus at the counter electrode, c) growth of CF phase due to the charged species -- of intrinsic in nature and of that which are -- supplied from the active electrode, and d) eventually electro-chemical process driven breaking of CF due to the application of electric bias of opposite polarity (in case of bipolar ECM memory cells). In various studies, the rupture of CF phase with the application of reduced magnitude and same polarity of applied voltage as compared to forming voltage have also been reported. Such switching memory devices are called unipolar ECM memory cells. In this paper, we consider the case of bipolar ECM memory cells.    

Computational approaches have been utilized previously to understand the \emph{I-V} response in the memristor systems. Such implementation are proved to be useful in obtaining the physical insight about the inner workings of memristor devices which are difficult to attain using experimental observations. For example, researchers~\cite{MenzelNanoscale, MenzelJAP, MenzelPCCP} have applied 1D physics based models with the information about electron transfer and ionic transport to understand the electric response in ECM devices as a function of filament-tip/active electrode separation distance. At atomic scale, researchers~\cite{Pan2010, Pan2011, Menzel2015, Dirkmann2017} have implemented Kinetic Monte Carlo method to validate the electrical responses -- driven by various materials and processing parameters -- in system specific experimental observations. In several such occasion the limitation associated with the 1D description of the filament/memresistive system~\cite{MenzelNanoscale} and the implementation of phenomenological probabilistic functions~\cite{Dirkmann2015,Li2017} to describe the electric field driven charge migration pose a situation where important physics of the problem can be missed out. To circumvent such limitation, we described the electric field driven filament formation using thermodynamically consistent phase field model~\cite{Roy2020}. The proposed phase field approach has proved to be sufficient to describe the electric field driven forming of conducting filament and to validate the all kind of experimental observations for any arbitrary memristor systems reported in the published articles. In this paper, we move forward and have used the similar phase field approach with the required modifications to understand the variability and reliability issues observed during the endurance cycle. We also could derive the useful understanding of device failure during the application of voltage sweeping cycle in endurance studies.

As we have already discussed, the phase field model used in this paper is similar to the one we have described in our previous publication~\cite{Roy2020}. Here, we provide a brief understanding of the main differences adopted in this study to simulate the endurance behaviour of ECM based ReRAM devices. In our previous paper, the release of cationic species has been realized using the constant flux as well as in cases using the Dirichlet boundary condition. On contrary, in this study we have updated the description of boundary condition to electric field modulated release of active cation to mimic the experimental observation reported in the literature~\cite{You2016,Kim2017ACS,Choi2021}. The details of the initial nucleus of the filament phase have been calculated prior to the simulation from the description of Gibbs free energy of the system and the nucleus has been implanted at the counter electrode in our previously published study~\cite{Roy2020}. In this study, as a precaution of maintaining the realistic situations, we have started our simulation without any nucleated filament phase, rather we wait for the migration and accumulation of cationic species at the counter electrode due to applied electric voltage, and have calculated the driving force for nucleation during the initial stages. Afterwards, while the driving force becomes sufficient to nucleate the filament phase, a seed of CF has been implanted at the counter electrode. The detail of the adopted phase field model and simulation conditions are discussed in the next sections.  
     
\section{Phase field modelling}
The adopted Kim-Kim-Suzuki (KKS) phase field model~\cite{Kim1998,Kim1999} for electrochemical system~\cite{Shibuta2006,Shibuta2007,Sherman2017,Roy2020} could elegantly capture the thermodynamics of the participating phases and the electric potential driven migration of charged species contributing to the kinetics of the system. We consider the following total free energy functional -- per molar volume -- of the system to maintain the thermodynamic consistency of the model: 
\begin{equation}
F = \int_V \left(f_{bulk} + \frac{\kappa_{\varphi}^2}{2} |\nabla \varphi|^2 + \frac{1}{2}\rho \Phi \right) dV. \label{E:totF}
\end{equation} 
$f_{bulk}$ denotes the bulk free energy density of the system. The gradient free energy density originating from the variation of a phase dependent variable called phase field ($\varphi$) is denoted by $ \frac{\kappa_{\varphi}^2}{2} |\nabla \varphi|^2$. The electrostatic energy density originating from the charge density ($\rho$) and the applied electric potential $\Phi$ is denoted by $\frac{1}{2}\rho \Phi$. 

The participating phases -- consisting of $\alpha$ memresistive matrix phase and the $\beta$ CF phase -- are distinguished in terms of phase field variable ($\varphi$) and composition ($C$). We choose $\varphi$ to be 0 and 1 inside the $\alpha$ and $\beta$ phases respectively. $\varphi$ varies smoothly from 1 to 0 at the filament/memresistive interface. The composition of the cation ($M^+$) rich $\alpha$ matrix and the metallic filament $\beta$ phase are denoted by $C_{M^+}^{\alpha}$ and $C_{M}^{\beta}$ respectively. The above mentioned phase dependent variables $\varphi$ and $C$ help us in writing the bulk free energy density description of the system as follows,
\begin{eqnarray}
f_{bulk} &=& (1 - p(\varphi)) f^{\alpha}(C^{\alpha}_{M^+}) + p(\varphi) f^{\beta}(C^{\beta}_{M}) \notag\\
&+& W g(\varphi). \label{E:fbulk}
\end{eqnarray}
We use ideal solution model to express the bulk free energy densities of the $\alpha$ and the $\beta$ phases as,
\begin{eqnarray}
f^{\alpha} &=& \left(G_{B}^{\alpha} + \frac{RT}{V_m} \ln \frac{1-\ ^e C^{\alpha}}{\ ^e C^{\alpha}}\right) C_{M^+}^{\alpha} \notag \\
&+& G_{B}^{\alpha}\left(1- C_{M^+}^{\alpha}\right) \notag \\
 &+& \frac{RT}{V_m}\Big[ C_{M^+}^{\alpha} \ln C_{M^+}^{\alpha}, \notag \\
 &+& (1 - C_{M^+}^{\alpha}) \ln (1 - C_{M^+}^{\alpha})  \Big] \label{E:bulkFa}
 \end{eqnarray}
 and
\begin{eqnarray}
f^{\beta} &=& \left(G_{B}^{\beta} + \frac{RT}{V_m} \ln \frac{1-\ ^e C^{\beta}}{\ ^e C^{\beta}}\right) C_{M}^{\beta} \notag \\
&+& G_{B}^{\beta}\left(1- C_{M}^{\beta}\right) \notag \\
 &+& \frac{RT}{V_m}\Big[ C_{M}^{\beta} \ln C_{M}^{\beta} \notag \\
 &+& (1 - C_{M}^{\beta}) \ln (1 - C_{M}^{\beta})  \Big] \label{E:bulkFb}
\end{eqnarray}
respectively. $^eC^{\alpha} (= 0.5)$ and $^eC^{\beta} (=0.98)$ are the assumed equilibrium concentration of $\alpha$ and $\beta$ phases. Here $R$ is the gas constant. $T = 300$ K and $V_m=1.4517\times10^{-5}$ m$^3$ mol$^{-1}$ are the temperature and the molar volume of the system. Also the $\alpha$ and $\beta$ phases are separated by a double--well potential $g(\varphi) = \varphi^2\left(1-\varphi\right)^2$ and the corresponding energy barrier of $W$ J m$^{-3}$. The details of the parametric values used in the description of the energetic of the system are mentioned in our previous publication~\cite{Roy2020}.

The variation of total free energy density $F$ (see Eq.~\ref{E:totF}) with respect to $\varphi$ is used to write the temporal evolution of $\varphi$,
\begin{eqnarray}
\frac{\partial \varphi}{\partial t} &=& M_{\varphi} \Bigg[ \kappa_{\varphi}^2 \nabla \varphi + \frac{d p(\varphi)}{d \varphi} \frac{R T}{V_m} \Big[f_{C_{M^+}}^{\alpha} - f_{C_{M}}^{\beta} \notag \\
&-& \frac{\partial f_{C_{M^+}}^{\alpha}}{\partial C_{M^+}} (C_{M^+} - C_{M})\Big] \notag \\
&-& W \frac{d g(\varphi)}{d \varphi}\Bigg]. \label{E:PF}
\end{eqnarray}

The diffusional flux provides the temporal evolution of ionic composition in the memresistive phase as follows,
\begin{eqnarray}
\frac{\partial C_{M^+}}{\partial t} = &\nabla& \Bigg[ \frac{D(\varphi)}{f_{cc}} \notag \\
&\nabla& \left( f_{C_{M^+}}^{\beta} + \frac{z_{M^+}F \Phi}{2  V_m}\right) \Bigg].  \label{E:Vcon} 
\end{eqnarray}
The phase field mobility is denoted by $M_{\varphi}$ and the diffusivity of the ionic charges are expressed as $D(\varphi) (=D_{\beta} p(\varphi) + D_{\alpha} (1-p(\varphi)))$. $D_{\alpha}$ ($= 3.0\times 10^{-18}$ m$^2$ s$^{-1}$) and $D_{\beta}$ ($= 3.0\times 10^{-20}$ m$^2$ s$^{-1}$) are the diffusivities of $M^{+}$ inside the $\alpha$ and $\beta$ phases respectively. By demanding the charge neutrality~\cite{Roy2020}, we obtain the distribution of electric potential due to the applied electric bias by solving the following expression,
\begin{eqnarray}
\nabla.\Big[ \varsigma(\varphi)\nabla \Phi\Big] &=& 0. \label{E:phi}
\end{eqnarray}
Here $\varsigma(\varphi) = \varsigma_{\beta} p(\varphi) + \varsigma_{\alpha} (1-p(\varphi))$ is the phase dependent ($\varsigma_{\alpha}$ for the $\alpha$ phase and $\varsigma_{\beta}$ for the $\beta$ phase) conductivity of the system.  

\subsection{Boundary condition to mimic the ionization process of the active electrode}
The process of electric field dependent ionization of charged particle at the active electrode is incorporated in the phase field model. For this purpose, we assume that the separation distance between the active electrode and the growing CF tip induces enhancement in the strength of developed electric field near the active electrode~\cite{Yang2014,You2016,Shin2016,Kim2017ACS,Sun2017,Cheng2019,Choi2021}. This assumption is quite straight forward to consider as the decreasing separation distance increases the strength of electric field considerably near the active electrode. This in turn influences the ionization process to release the cation species in the memresistive layer. 

To realize the ionization of active electrode though the boundary condition, we propose the following electric field dependent Dirichlet boundary condition at the active electrode,
\begin{equation}
C_{Dirichlet}(\Phi) = C_0 \exp\left(-\delta\frac{E_{force}}{RT/V_m}\right). \label{E:BC1}
\end{equation}
Here $C_0$ is the composition for Dirichlet boundary condition independent of electric field. $E_{force}$ mimics the electrostatic force originating due to the electric field. With the approaching metal filament, the enhancement of electric field initiates the ionization of active electrode and release of metal cation. To describe the release of such cation as a result of ionization of the active electrode, we consider the energy to be $\delta E_{force}$. Here $\delta$ is a simulation constant with a dimension of length and corresponds the atomic separation distance. We choose $\delta$ to be $0.14517$ nm. 
\section{Numerical implementation}
We use second order accurate finite difference method to solve the temporal evolution of phase field (Eq.~\ref{E:PF}) and cationic composition (Eq.~\ref{E:Vcon}). The distribution of electric potential is obtained by solving Eq.~\ref{E:phi} using the Gauss-Seidel method. In our simulation, the diffusivity in Eq.~\ref{E:Vcon} and the conductivity in Eq.~\ref{E:phi} varies with the participating phases; CF phase ($\varphi = 1.0$) and memresistive phase ($\varphi=0.0$). We employ sixth order Lagrange interpolation formula to resolve the phase dependent diffusivity and conductivity at the mid-point of two consecutive grids while implementing the finite difference scheme at the interface. In our simulation domain, the top and and the bottom boundary represent the active and the counter electrode respectively. We employ adiabatic boundary condition at the bottom boundary and the Dirichlet boundary condition at the top boundary as has been discussed in the previous section. The periodic boundary condition is implemented at the left and right boundaries. The numerical simulations have been carried out using CUDA-C for GPU parallelized computation on Nvidia Titan Xp graphics card. Following expression is used to compute the electric current originating in the system,
\begin{equation}
I = \int_S \varsigma \vec{E}.\vec{n}dS \label{E:elecCurrent}
\end{equation}
$\vec{E}$ represents the electric field calculated at the interface from the spacial distribution of $\Phi$. The conductivity and the interface normal are denoted using $\varsigma$ and $\vec{n}$ respectively. The surface element at the conducting filament / memresistive interface is realized using $dS$. At this point we also want to mention that the voltage at which an abrupt increase in measured electric current has been observed is termed as forming (in case of first such occurrence) or SET (in case of consecutive occurrences). Similarly, the voltage at which an abrupt decrease in electric current occurs is termed as RESET voltage. 

\section{Results and discussions}

\begin{figure*}[htbp]
\centering
\includegraphics[width=6.3in]{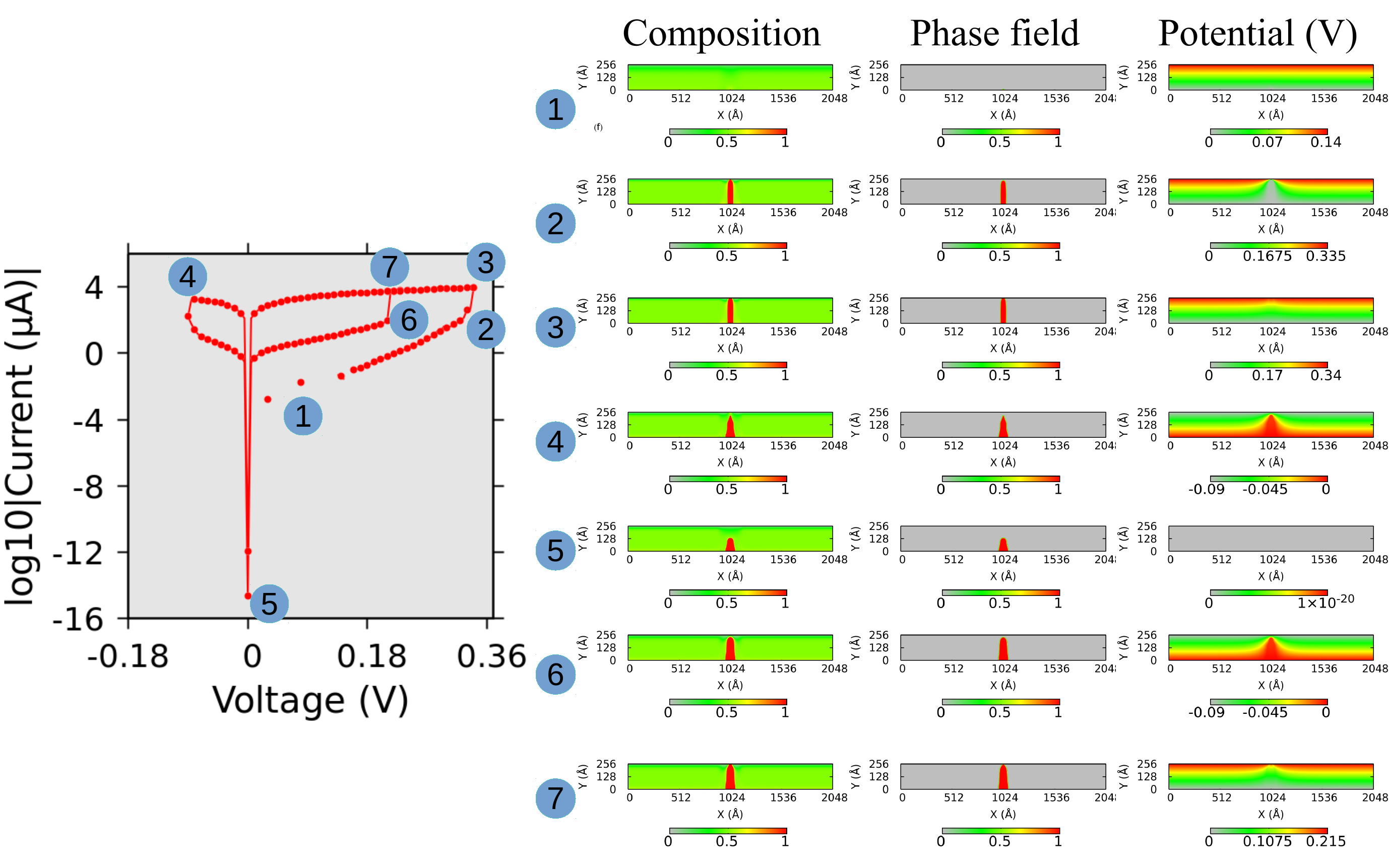}
\caption{Morphological evolution of CF phase during the application of voltage sweeping rate (0.01 V/s) and the corresponding \emph{I-V} curve. At stage 1, CF phase nucleates in the memresistive phase. Stage 2, CF phase grows sufficiently due to the applied voltage. Forming of CF phase completes and the measured current in the system changes abruptly at stage 3. By reversing the sweeping voltage direction filament breaks (RESET initiates) at stage 4 for negative applied voltage. Stage 5 describes the morphology of CF phase while reversing the voltage sweeping rate to +0.01 V/s at 0.0 V. At stage 6, with the increasing magnitude of voltage sweeping rate filament again starts to grow. At stage 7, the filament short-circuit the counter and active electrodes and the measured current increases abruptly leading to SET process}\label{F:mrphEvol}
\end{figure*}

\begin{figure}[htbp]
\centering
\includegraphics[height=2.0in]{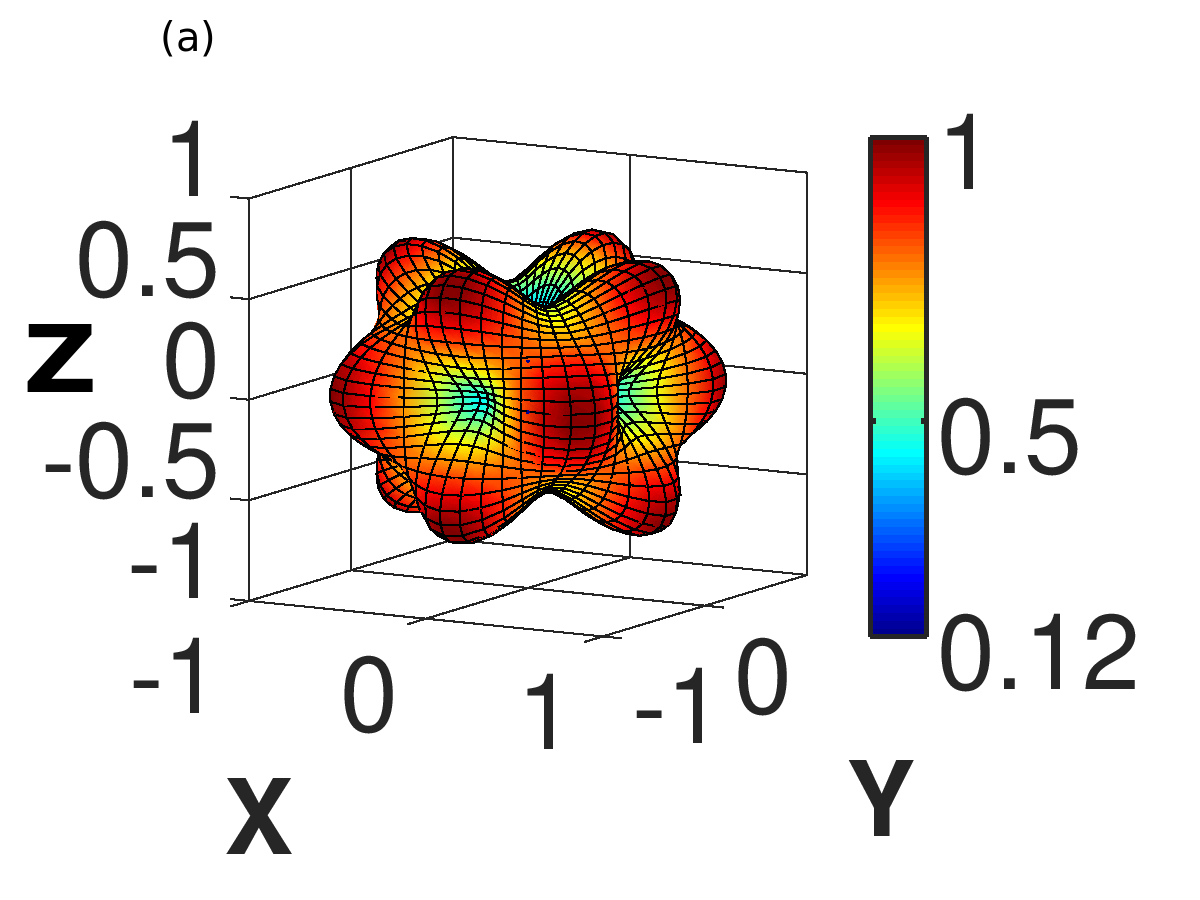}
\includegraphics[height=2.0in]{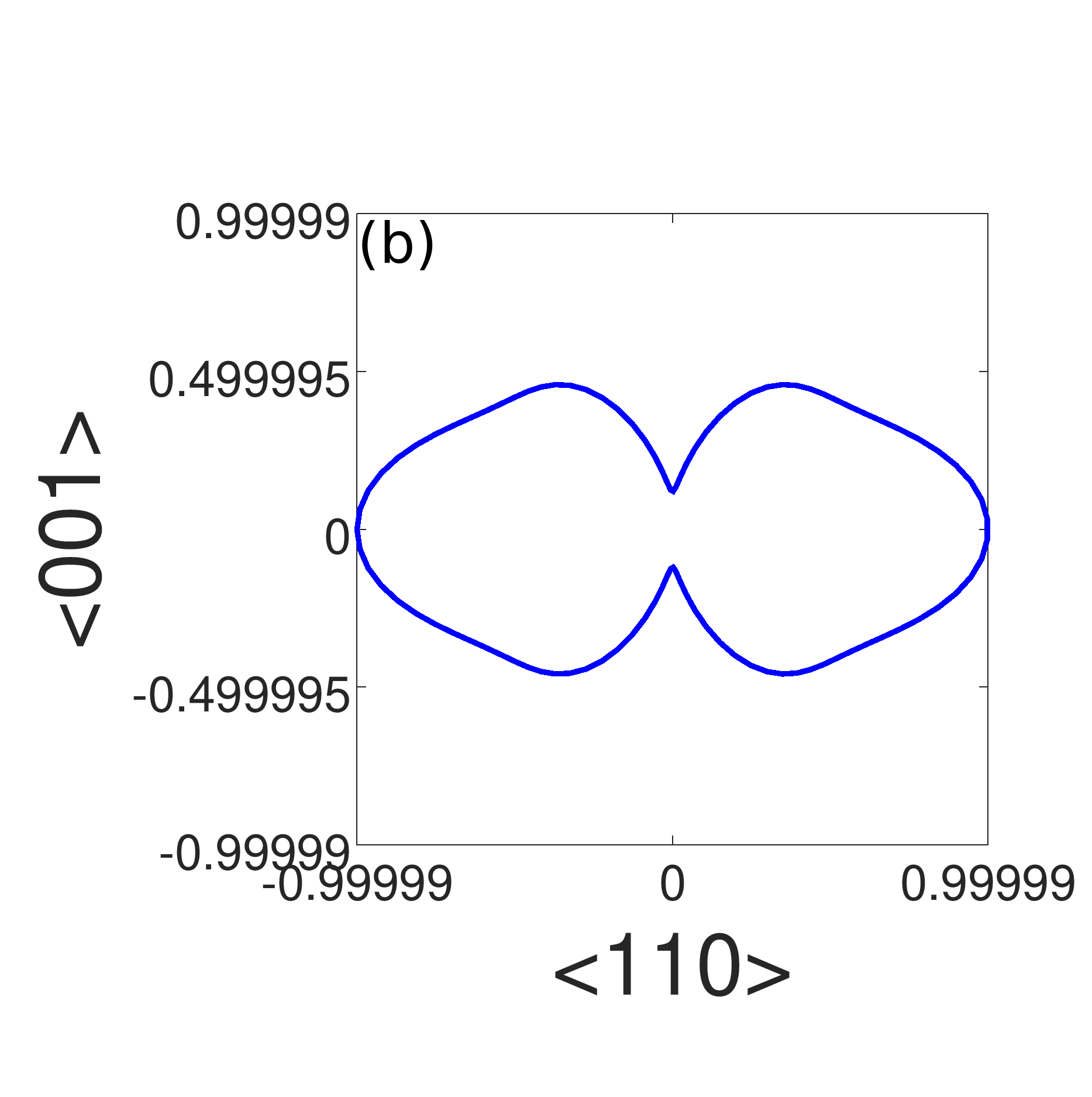}
\caption{(a) 3-D polar plot of anisotropic growth kinetics factor presented in Eq.~\ref{E:anisoM} with $n_1 = 0.44444$, $n_2 = 0.12$, $n_3 = 3.5555$, $n_4 = 4.4444$, $n_5 = 2.6666$, and $n_6 = -4.0$. (b) 2-D section of polar plot on the simulation crystallographic direction; low growth kinetics along the parallel direction of counter electrode and high growth kinetics towards the direction of active electrode.}\label{F:growAn}
\end{figure}

In Fig.~\ref{F:mrphEvol}(a) we show the {\emph{I-V}} curve obtained from our simulation due to the application of sweeping voltage at a rate of $0.01$ V/s. As evident from the figure, our model can be used to simulate the forming (refer to our article~\cite{Roy2020} for forming process related details), SET, and RESET processes elegantly. To understand the morphological changes of CF during the {\emph{I-V}} response, we have presented the corresponding micrograph of the CF at different stages of the voltage sweeping cycle in Fig.~\ref{F:mrphEvol} as well. Application of electric field -- which controls the release of charged cationic species and migration towards the counter electrode -- by sweeping the biasing voltage in the memresistive system, modulate the growth of the filament. We want to note that to restrict the uncontrolled broadening of CF along the counter electrode (where the charged species are abundant due to the accumulation driven by applied field), we have incorporated the anisotropy in growth kinetics. The evidence of such preferred growth directions induced by the the CF phase has been explored in HfO$_2$ system through the help of ab initio calculations~\cite{Duncan2016}. The anisotropy has been introduced through the orientation dependent diffusivity (of the charged species in the memresistive layer) and the phase field mobility of the system. We have assumed that the growth is enhanced along the direction of active electrode (towards which the filament is growing) and is reduced along parallel to the counter electrode direction. For such direction dependent anisotropy, we have considered tetragonal crystalline symmetry. The anisotropic polynomial ($P(n_x,n_y,n_z)$) has been developed by Roy and Gururajan~\cite{RoyACS2021}, and it takes the form for anisotropic kinetics factor as follows:
\begin{eqnarray}
P(n_x,n_y,n_z) &=& n_1 (n_x^6+n_y^6) + n_2 n_z^6 \notag \\
&+& n_3 n_x^2 n_y^2 (n_x^2 + n_y^2) \notag \\
&+& n_4 n_z^2(n_x^4+n_y^4) + n_5 n_z^4(n_x^2+n_y^2) \notag \\
&+& n_6 n_x^2 n_y^2 n_z^2 \label{E:anisoM}
\end{eqnarray}
The 3-D polar plot and the 2-D section of the anisotropic polynomial, chosen for the simulation frame of reference are presented in Fig.~\ref{F:growAn}(a) and (b) respectively. As intended, we can see that the magnitude of the polynomial is higher along the direction of active electrode than that of the counter electrode. Such choice of anisotropy in growth kinetics restrict the uncontrolled broadening of CF along counter electrode direction. At this point, we want to note that to save the computational cost, all our data related to electric responses are simulated on 2-D domain. However, our model can be easily extended to 3-D system and are shown in our previous publication~\cite{Roy2020}, which uses a phase field model to understand the physio-chemical-electrical stimuli driven formation mechanism of CF in memresistive system. 

We also want to note that during the operation condition, it is possible that the misfit strain originating due to the mismatch between lattice parameters of CF phase and memresistive phase give rise to the elastic energy contribution. However, in our model we consider that both the existing phases pose similar molar volume and the participating interface is incoherent in nature. Also, as we already have an existing iterative solver for the calculation of electric potential, it would be computationally intensive if we add an other iterative solver for elastic energy calculation. So, we believe that our approach of controlling the CF thickness is adequate in the current scope of our numerical simulation.

\begin{figure}[htbp]
\centering
\includegraphics[width=2.2in]{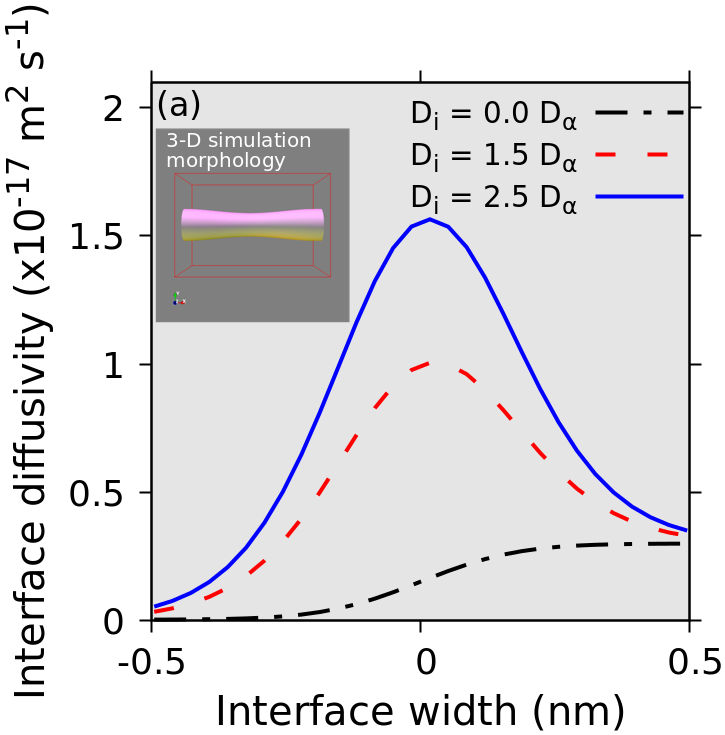}
\includegraphics[width=2.2in]{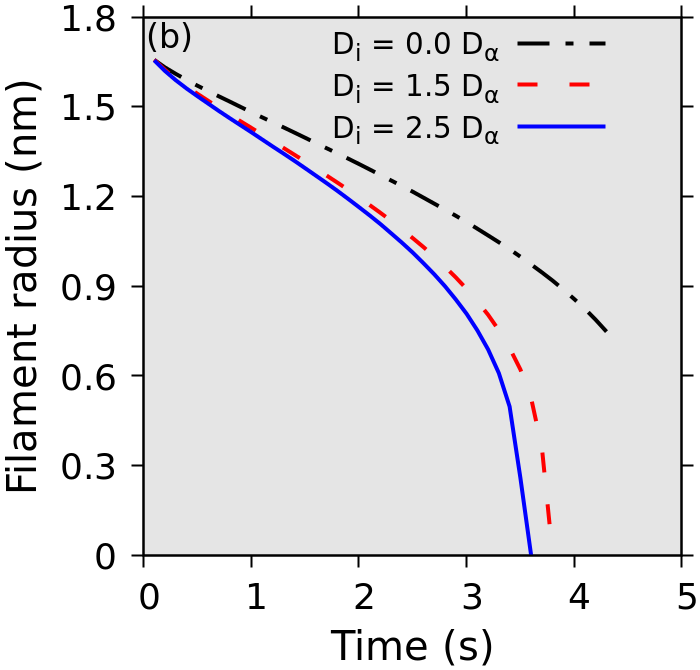}
\caption{(a) variation of surface diffusivity at the filament--memresistive interface. A representative 3-D simulation morphology used to extract the effect of surface diffusivity is shown at the inset. (b) variation of filament radius as a function of time for varying degree of surface diffusivity.}\label{F:surfDif}
\end{figure}

In addition to the anisotropy in diffusivity of the charged species at the memresistive layer, the cationic diffusivity at the interface has also been enhanced. The chemical driving force responsible for the charge-migration due to the applied electric field becomes dominant due to such enhancement of the diffusivity at the interface. The morphological changes of CF phase as a consequence of dominant surface diffusion of charged particle has also realized using molecular dynamics simulation study on Ag/Silk-Ag nanowires memresistive system~\cite{Wang2019}. We as well find that in the presence of physical factors responsible to modulate the shape and charge-migration are essential in capturing the morphology controlled electric response of the ReRAM system. Enhanced diffusion accelerate the charge-migration from the interface. The TEM observations on CF and memresistive interface also explain the influence of surface diffusivity on the morphological changes in the ReRAM devices~\cite{Yang2014, Chae2017}. Such influence leads to the electric field controlled breaking of the CF. The distribution of the enhanced diffusivity at the interface and its effect on the filament thickness have been shown in Fig.~\ref{F:surfDif}(a) and (b) respectively. We extract the enhanced diffusivity effect on the filament breaking from a set of 3-D simulations, where a notch is introduced at the middle of the filament. The 3-D morphology used in the simulation has been presented in the inset of Fig.~\ref{F:surfDif}(a). We choose 2-D simulation domain to understand the effect of voltage sweeping cycle on the morphological evolution of the CF. In this purpose the value of diffusivity at the interface has been enhance by a factor of 7.0 as compared to bulk cation diffusivity in the memresistive phase ($D_{\alpha} = 3.0\times10^{-18}$ m$^{2}$.s$^{-1}$). 

In our previous study to understand the effect of physio-chemical factors on the growth of CF and the resulting electric response, we started our simulation with the already nucleated seed of CF phase in the memresistive layer~\cite{Roy2020}. To mimic the possible experimental condition and to make the study more general, we consider the classical nucleation theory to implant the initial seed of CF phase. We consider the diffusion of cationic species in the memresistive layer toward the counter electrode due to the applied electric voltage at the active electrode. The accumulation of the cationic concentration at the counter electrode increases the nucleation rate ($\Re$), which is a function of number of nucleation sites ($N_S$), attachment rate of new molecules with the nuclei ($j$), Zeldovich factor ($Z$), and the driving force for nucleation ($\Delta G^*$). Due to the increased cationic concentration at the counter electrode, $\Delta G^*$ becomes significant and increases the availability of new nucleation sites. Following the classical nucleation theory, $N_S$ can be written as,
\begin{eqnarray}
N_{S} &=& \frac{\Re}{j Z} \exp \left(\frac{\Delta G^*}{\frac{RT}{V_m}}\right). \label{E:Ns}
\end{eqnarray} 
By identifying the $\Re$, we can control the number of CF phase nucleus at the counter electrode, which can be implanted at random locations. In the present scope of study, we choose the $\Re$ to be 0.5. The cut off for $jZ$ is set at 0.6. Such choice of parameters helped us to restrict the number of implanted nucleus to 1. However, we can adjust the simulation parameters in case we are interested in the situation of multiple CF growing simultaneously. The radius of the implanted nucleus is set at 2.0 nm and is consistent with the size of critical nucleus required for nucleation~\cite{Roy2020}. As we deal with only one critical nucleus seed, we place the seed at the middle of the counter electrode. 

\begin{figure*}[htbp]
\centering
\includegraphics[width=1.5in]{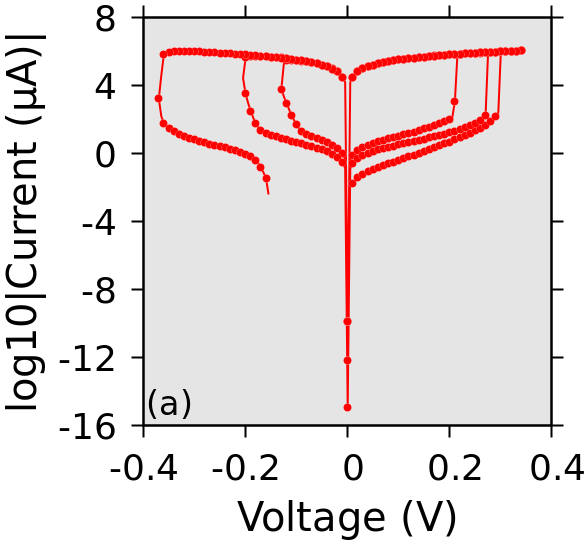}
\includegraphics[width=1.5in]{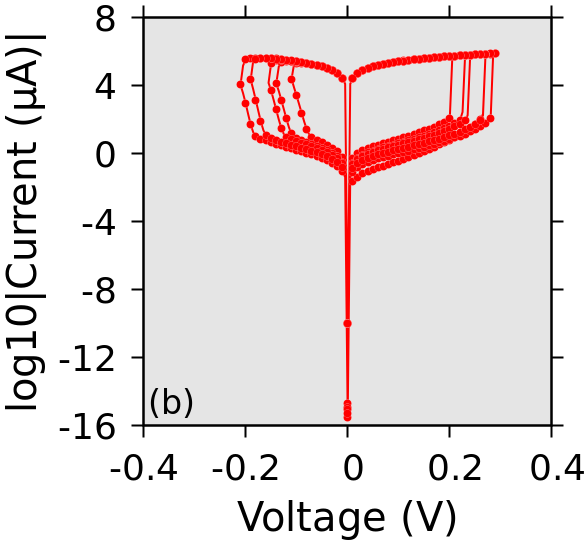}\\
\includegraphics[width=1.5in]{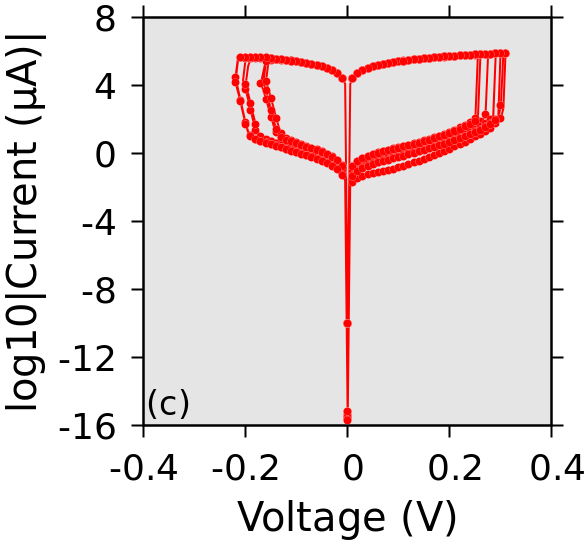}
\includegraphics[width=1.5in]{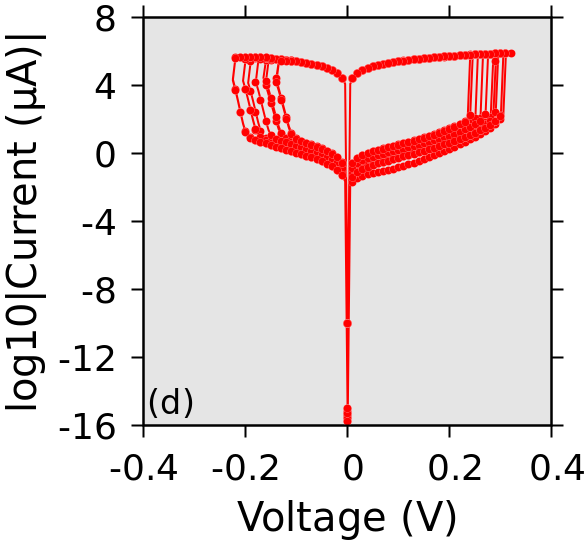}\\
\includegraphics[width=1.5in]{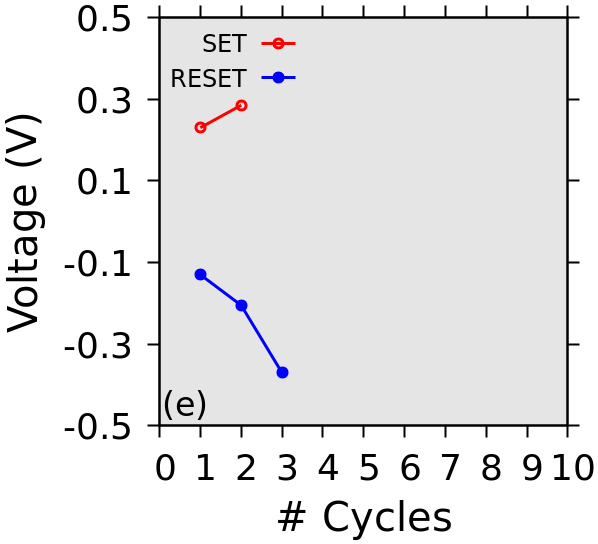}
\includegraphics[width=1.5in]{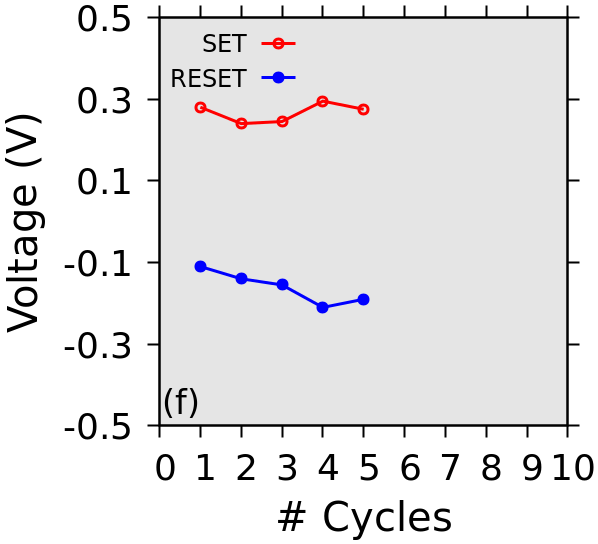}\\
\includegraphics[width=1.5in]{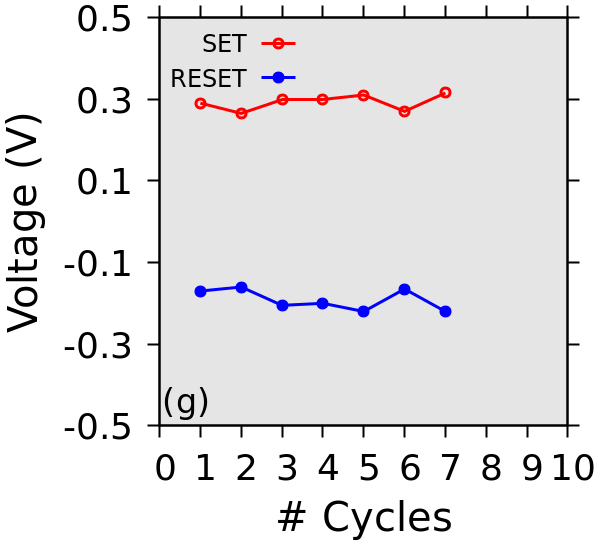}
\includegraphics[width=1.5in]{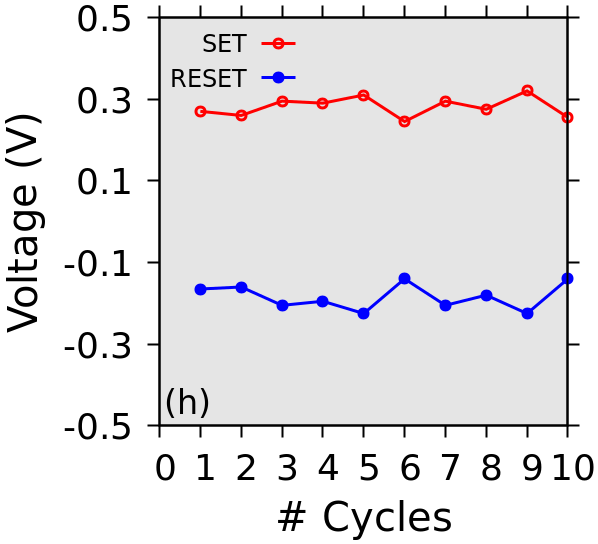}
\caption{{\emph{I-V}} curves obtained using a cationic supersaturation of $0.55$, $0.51$, $0.501$, and $0.5001$ are shown in (a), (b), (c), and (d) respectively. The corresponding SET and RESET voltage variations with the number of sweeping cycles are shown in (e), (f), (g), and (h) respectively. }\label{F:intriDef}
\end{figure*}

We fix the details of the anisotropic growth kinetics (in terms of cation diffusivity and phase field mobility), the enhanced surface diffusivity, and the nucleation of filament seeds discussed so far for the rest of our study. In order to understand the effect of intrinsic cationic concentration inside the memresistive layer of the ReRAM devices, we have varied the supersaturation of cationic species. From the obtained \emph{I-V} curves, as shown in Fig.~\ref{F:intriDef}, we understand that the increasing intrinsic defects of charged species are not favourable for the operation of ReRAM devices. We find that the system with lowest intrinsic defects density reaches to 10 sweeping cycles mark. We set the number of sweeping cycle to 10 in order to choose the materials and processing parameters of the simulation as a benchmark of comparatively better performing device.    

\begin{figure*}[htbp] 
\centering
\includegraphics[width=1.5in]{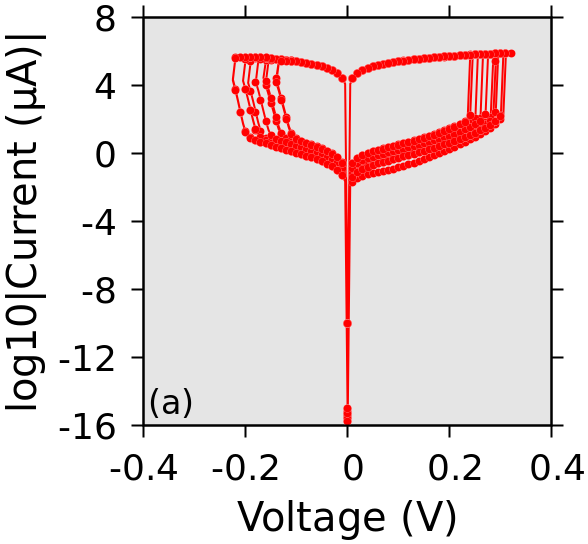}
\includegraphics[width=1.5in]{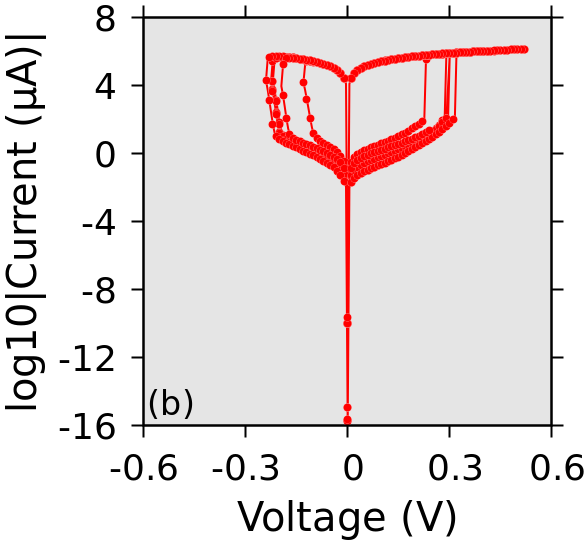}\\
\includegraphics[width=1.5in]{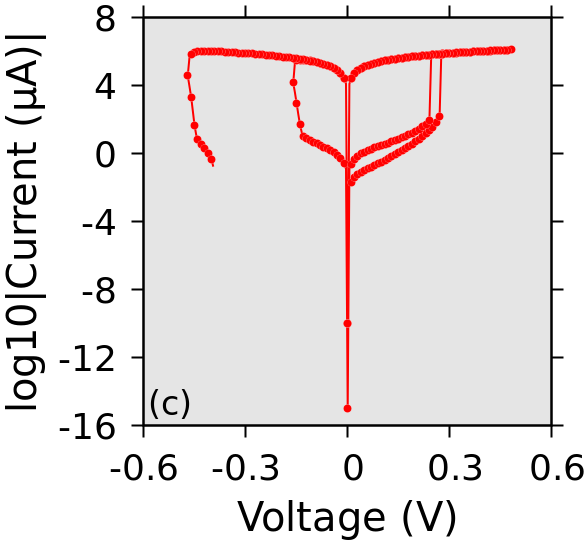}
\includegraphics[width=1.5in]{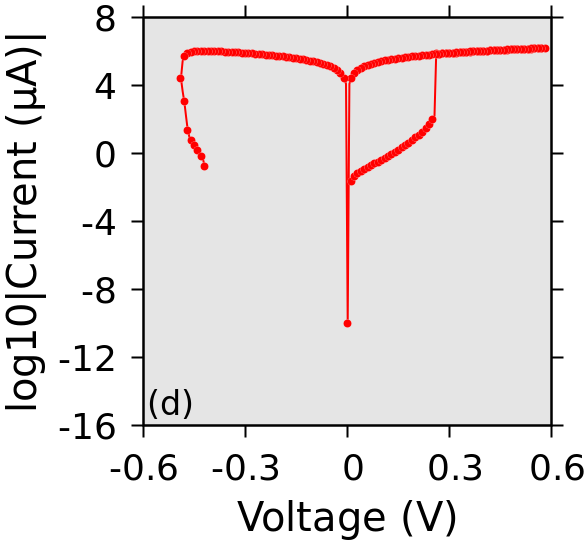}\\
\includegraphics[width=1.5in]{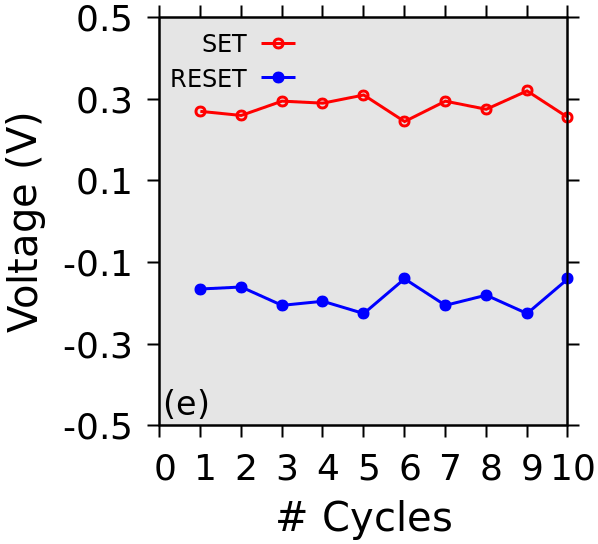}
\includegraphics[width=1.5in]{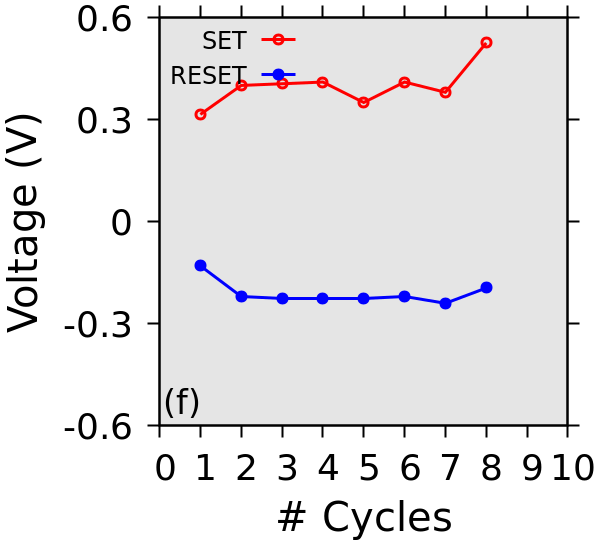}\\
\includegraphics[width=1.5in]{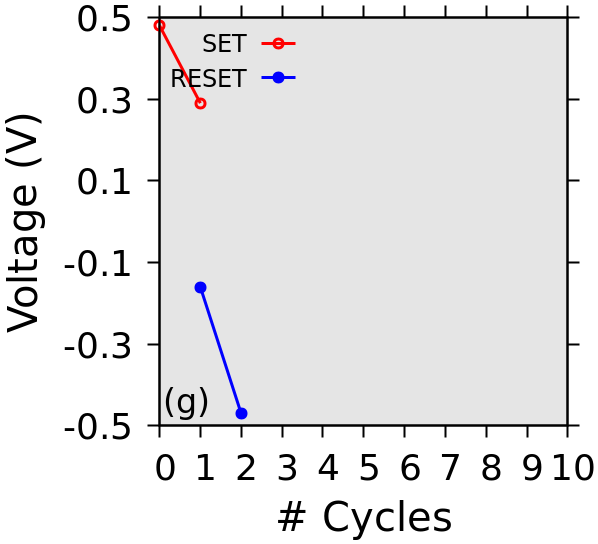}
\includegraphics[width=1.5in]{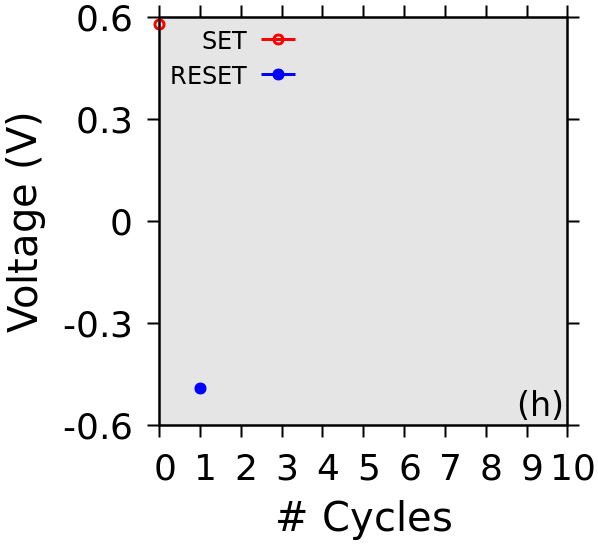}
\caption{{\emph{I-V}} curves obtained for a choice of filament/memresistive interfacial energy (J.m$^{-2}$) of $0.05$, $0.0375$, $0.025$, and $0.0125$ are shown in (a), (b), (c), and (d) respectively. The corresponding SET and RESET voltage variations with the number of sweeping cycles are shown in (e), (f), (g), and (h) respectively.}\label{F:intfEng}
\end{figure*}

Once we identify the cationic supersaturation, favourable for the growth of CF and ideal for maintaining minimum sweeping cycles mark (which is 10 as discussed), we move towards understanding the parameters that control the length scales associated with the growth of the filament. We vary the filament/memresistive layer interfacial energy, which can withstand the predefined sweeping cycle mark as discussed before. In this process, we identify that the higher interfacial energy contribution leads to more stable filament growth and stable \emph{I-V} response, which can also be seen in Fig.~\ref{F:intfEng}.

\begin{figure*}[htbp]
\centering
\includegraphics[width=1.5in]{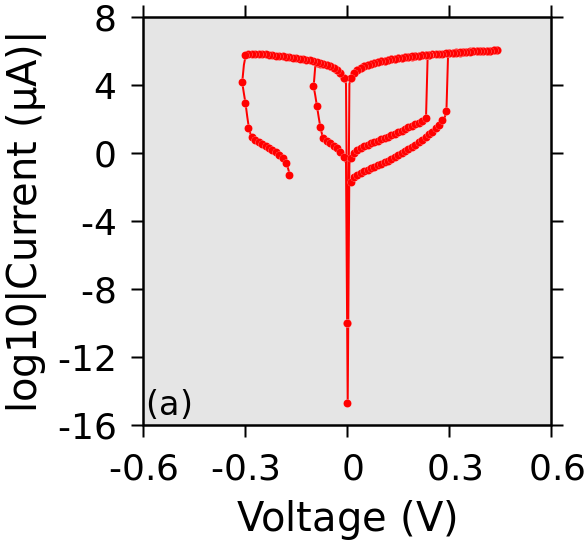}
\includegraphics[width=1.5in]{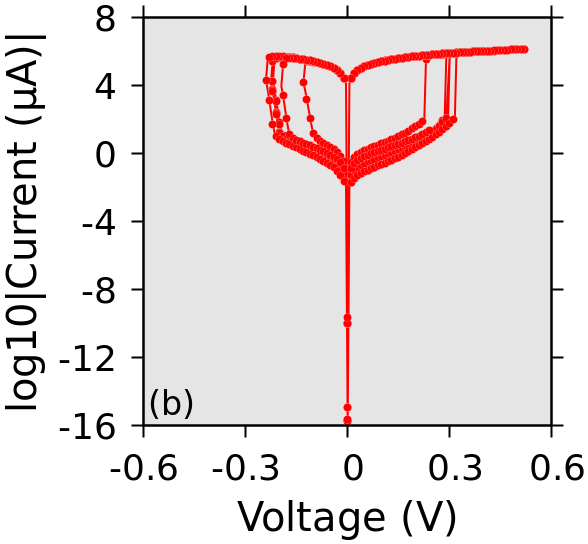}\\
\includegraphics[width=1.5in]{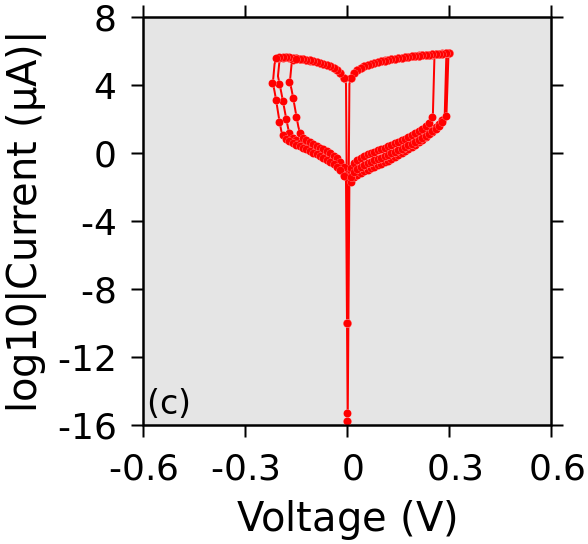}
\includegraphics[width=1.5in]{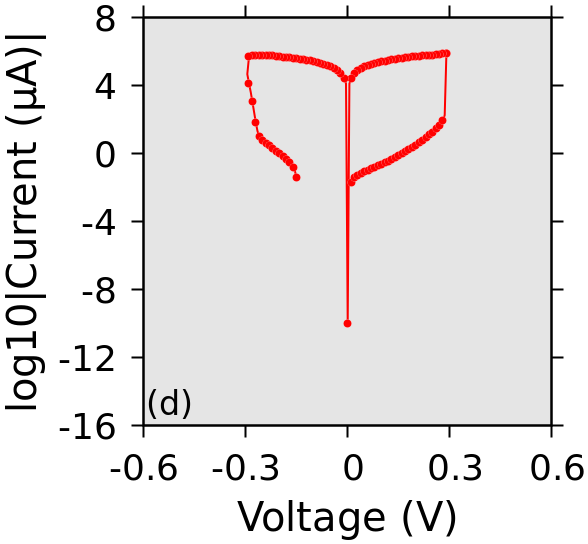}\\
\includegraphics[width=1.5in]{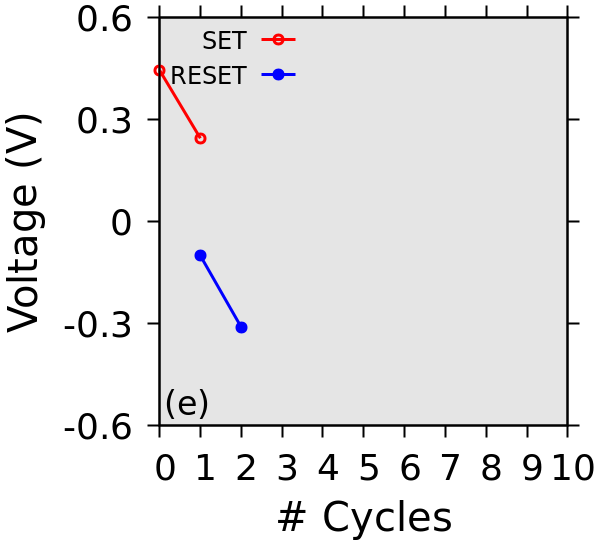}
\includegraphics[width=1.5in]{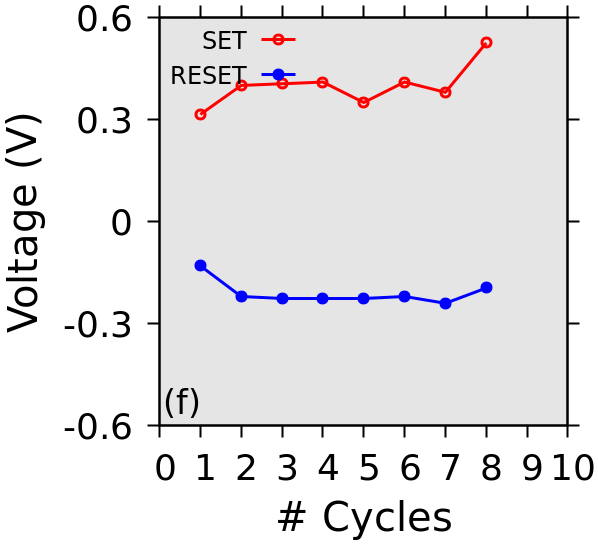}\\
\includegraphics[width=1.5in]{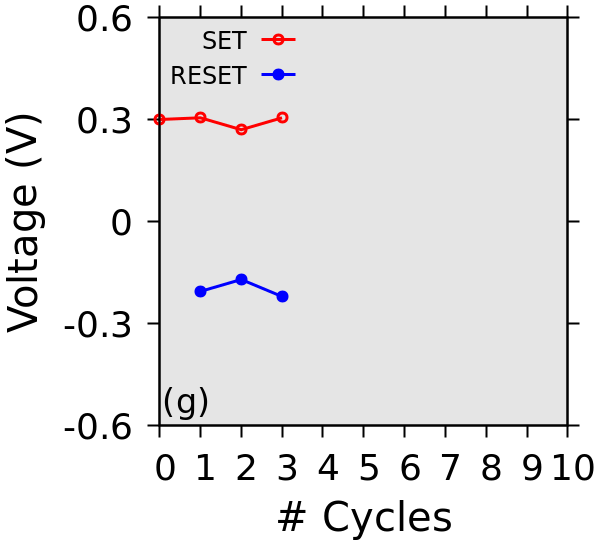}
\includegraphics[width=1.5in]{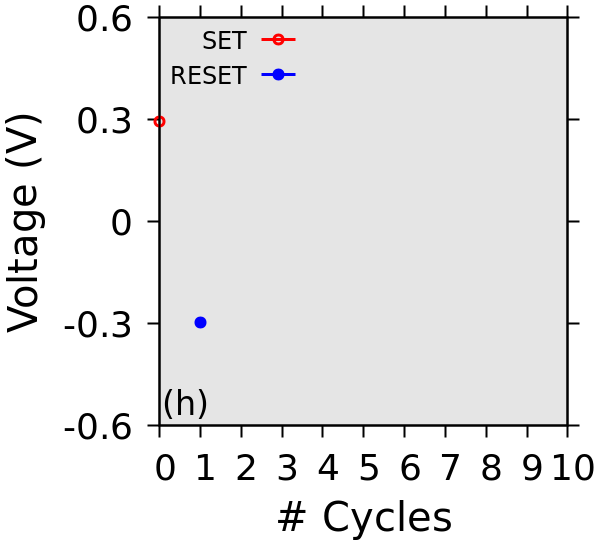}
\caption{Effect of cation species released from the active electrode (implemented using Dirichlet boundary condition) on the \emph{I-V} response and the variation of SET/RESET voltage with the sweeping cycles are presented in this figure. {\emph{I-V}} curves obtained using a cationic concentration (at the active electrode) of $0.05$, $0.10$, $0.15$, and $0.20$ are shown in (a), (b), (c), and (d) respectively. The corresponding SET and RESET voltage variations with the number of sweeping cycles are shown in (e), (f), (g), and (h) respectively.}\label{F:effBC}
\end{figure*}

The growth of the filament and the stability of \emph{I-V} response are also controlled by the choice of electrode materials. The top electrode or active electrode layer not only controls the applied bias in the memresistive layer, but is also responsible for the release of active charged species in the ECM based resistive switching devices. We use the magnitude of $C_0$ as expressed in Eq.~\ref{E:BC1} as the parameter that controls the release of cationic species in the memresistive layer. From Fig.~\ref{F:effBC}, we can see that $C_0 = 0.1$ is susceptible in maintaining fair number of voltage sweeping cycles. 

\begin{figure}[htbp]
\centering
\includegraphics[width=2.2in]{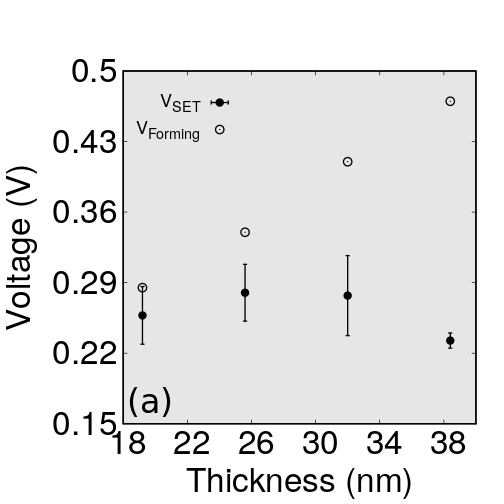}
\includegraphics[width=2.2in]{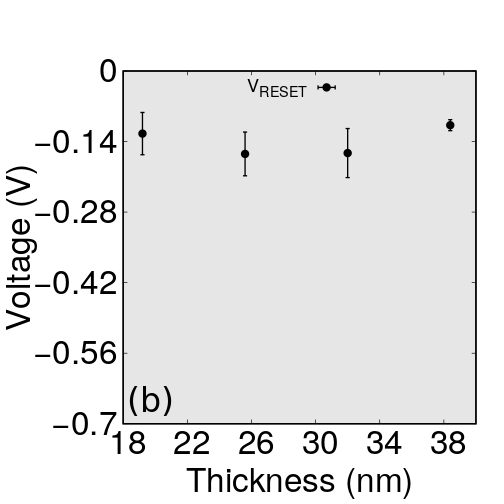}
\caption{The effect of device thickness on the (a) forming and SET voltage, and (b) RESET voltage. The data for SET and RESET voltages carry error bars as they are obtained from 10 cycle endurance simulations.}\label{F:effThick}
\end{figure}

The bipolar switching behaviour observed during the voltage sweeping cycles are similar to the cases of a broad range of memresistive systems as reported in the literature. As per the observation in various experimental literature, we find that the forming voltage is always greater than the SET voltage. As shown in Fig.~\ref{F:effThick}(a), the forming voltage increases almost linearly with the device thickness~\cite{Schindler2009,Tian2017,Yoo2015}. On the other hand the variation of SET and RESET voltages remain almost invariant with the change in device thickness~\cite{Bricalli2018,Bricalli2018_2}. Also, the absolute magnitude of SET voltage is always greater than that of the RESET voltage. The filament phase does not dissolve completely after the RESET process. Such remnant filament phase helps in growing the filament faster during the SET process as compared to the forming process. For this reason, the time to complete the filament formation remains relatively low as compared to forming process with the varying device thickness -- leading to low SET voltage as compared to the voltage required during the forming process. On the other hand, in Fig.~\ref{F:effThick}(b) we can see that the breaking of filament phase depends on the choice of the chemical system involved in the ReRAM devices and on the applied electric bias. For this reason, the change in RESET voltage with the device thickness does not show any specific trend.   

\begin{figure}[htbp]
\centering
\includegraphics[width=2.2in]{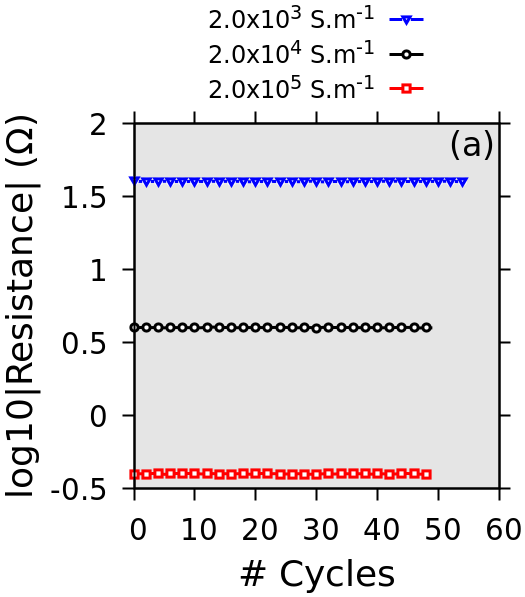}
\includegraphics[width=2.2in]{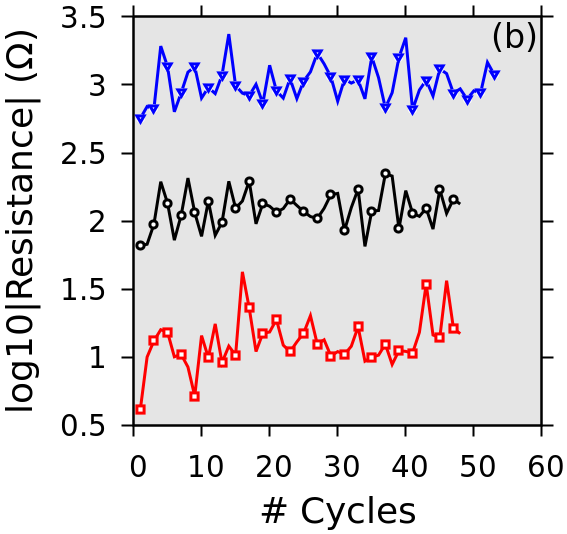}
\caption{Variation of (a) SET and (b) RESET resistances with the application of voltage sweeping cycles. The resistances are calculated from the {\emph{I-V}} curves for systems with varying CF phase conductivities; $\varsigma_{\beta} = 2.0\times10^3$ S.m$^{-1}$, $2.0\times10^4$ S.m$^{-1}$, and $2.0\times10^5$ S.m$^{-1}$.}\label{F:varResis}
\end{figure}

\begin{figure}[htbp]
\centering
\includegraphics[width=2.2in]{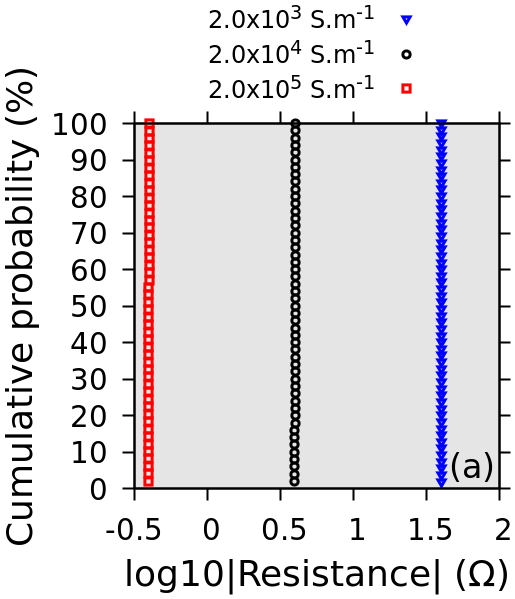}
\includegraphics[width=2.2in]{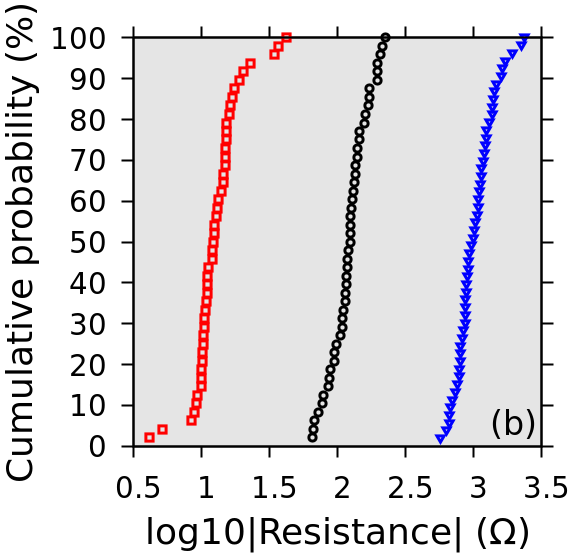}
\caption{Cumulative probability distribution of (a) SET and (b) RESET resistances calculated for data presented in Fig.~\ref{F:varResis} for $\varsigma_{\beta} = 2.0\times10^3$ S.m$^{-1}$, $2.0\times10^4$ S.m$^{-1}$, and $2.0\times10^5$ S.m$^{-1}$.}\label{F:varCDF}
\end{figure}

By varying the material parameters, we are able to identify the physio-chemical stimuli that control the stability of the electric response in the ReRAM devices. However, the variation in the current and voltage values during SET and RESET processes are visible in the {\emph{I-V}} responses we presented so far in this article. We try to understand the origin of the variability observed in the {\emph{I-V}} responses. We consider a model ReRAM device which is stable for multi-cycle voltage sweeping or endurance cycle. We choose varying magnitudes of conductivity of the CF phase and the conductivity of the memresistive phase is fixed at 2.0 S.m$^{-1}$. In Fig.~\ref{F:varResis}(a) and (b) the variation of resistivity as a function of endurance cycles has been shown for the LRS and HRS state respectively. To recreate the possible real experimental set up, we have also restricted the sweeping cycle to a specific sweeping bias range; -0.36 V to +0.36 V. In accordance with the experimental studies~\cite{Lee2012,Shin2016,Sun2017,Choi2021}, we find that the fluctuation in the magnitude of measured resistance -- from the electric responses during endurance -- for the LRS state is much low as compared to the HRS state. To understand spread of variation in the resistance state, we calculate the empirical cumulative probability of the resistances in the LRS and the HRS state and have shown in Fig.~\ref{F:varCDF}(a) and (b) respectively. As we can see, the spread in LRS resistivity are small as compared to HRS resistivity. The observation is consistent with that reported in the experimental study for a variety of ReRAM devices~\cite{Ambrosi2019,Nail2017,Bricalli2018}. However, the spread in HRS resistivity is less in case of a model ReRAM device where the ratio of conductivities between the CF and the memreistive phase is 10$^{4}$. Such small variation in empirical cumulative probability are desirable for reliable switching behaviour of the ReRAM devices. 

\begin{figure}[htbp]
\centering
\includegraphics[width=2.2in]{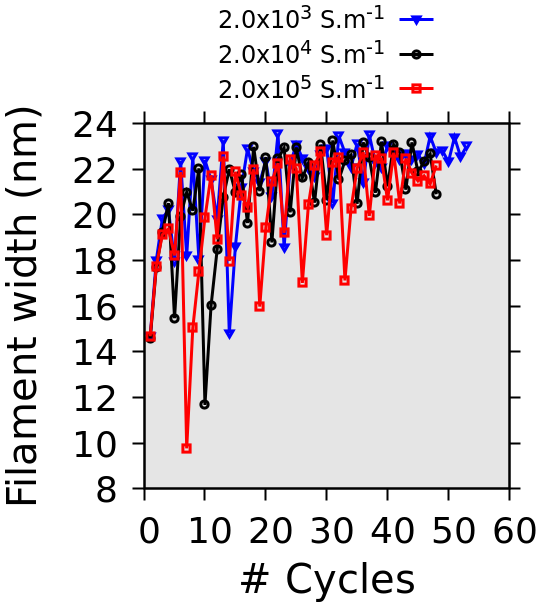}
\caption{Change in filament radius with the application of voltage sweeping cycles. The state of the resistance remains at LRS due to such broadening of CF and eventually leads to the device failure.}\label{F:varBroad}
\end{figure}

From the discussion so far, we understand that the presented phase field study can predict the model ReRAM system in terms of device stability and, variability and reliability in electrical responses elegantly. It is also possible to predict the probable reason of device failure with the repeating endurance cycles. In Fig.~\ref{F:varBroad}, we show the change in filament thickness with the endurance cycle. We find that irrespective of the choice of ReRAM system, the filament broadens as we increase the endurance cycles. Such broadening of filament thickness with the applied endurance cycle has serious consequence on the device operations. In case when the filament broadens considerably during the endurance cycle, it becomes ever struggling for the applied negative bias to bring the system back to HRS by breaking the filament in the RESET process. It results in ReRAM devices stuck at the LRS state and failed to achieve the HRS state, leading to device failure. In accordance with the experimental observations~\cite{Lv2015,Wang2016,Gong2018,Dutta2020,Bejtka2020}, our simulation study also point towards the broadening of CF as a dominant mechanism of device failure. At this point it is important to realise that the increasing RESET voltage -- leading to deteriorating device response capability and the possibility of device staying at LRS state indefinitely -- could mark the onset of imminent device failure. Also, we want to note that though we point towards the mechanism of possible device failure due to the broadening of filament, we have also discussed the way of avoiding it while discussing the stability of {\emph{I-V}} curves. In contrary to the study of variability and reliability issues of electrical responses, in case of studying the stability issue, we do not put any restriction in the sweeping bias limit. The polarity of sweeping rate has been reversed once the SET or RESET process has been achieved. However, in case when the device failed to maintain the minimum 10 endurance cycle marks, we find that the difference between SET and RESET voltage always increases, which ultimately failed to show the desirable endurance behaviour.

\section{Conclusions}
We recreate a broad spectrum of issues usually encountered during the operation of ReRAM devices in terms of stability, variability, and reliability of electrical responses during the multi-cycle endurance behavior. We predict the set of physio-chemical stimuli which control the electrical responses during the bipolar switching mechanism and have identified the key device parameters in terms of intrinsic cationic concentration, the parameters that control the release of cations from the active electrode, and the choice of filament system with appropriate interfacial energy and conductivity with respect to the memresistive phase. Understanding the issues which are usually encountered during ReRAM device operations using the implemented thermodynamically consistent phase field model are proved to be beneficial in obtaining desirable ReRAM systems in terms of materials parameters and operation conditions. We believe that our model system, when applied to identify a suitable ReRAM device material with desirable electrical responses, could prove to be effective in understanding and solving the key operation challenges.
\section*{Acknowledgements}
We acknowledge Creative Materials Discovery Program through the National Research Foundation of Korea (NRF) funded by the Ministry of Science, ICT and Future Planning (2016M3D1A1027666), and National Research Foundation of Korea (NRF) grant funded by the Korea government (MSIP) (NRF - 2019R1A2C1089593 and NRF - 2019M3A7B9072144) for funding support.
\section*{Data availability}
The data that support the findings of this study are available from the corresponding author upon reasonable request.

\end{document}